\documentclass[conference]{IEEEtran}
\usepackage{blindtext, graphicx}

\usepackage{graphicx}
\usepackage{comment}
\usepackage{adjustbox}
\usepackage{amsmath}
\usepackage{subcaption}

\usepackage{verbatim}

\begin{document}
%
\title{Powering Future Mobile Phones Through RF Energy Harvesting }
\author{\begin{tabular}{cc}
 &\textbf{ Ankush Jolly, Mansi Peer, Vivek Ashok Bohara, } \\
 & Wirocomm Research Group, Department of Electronics and Communication, IIIT-Delhi, India\\
& Email: ankush12128@iiitd.ac.in, mansip@iiitd.ac.in, vivek.b@iiitd.ac.in\\
 \end{tabular}
 } 
\maketitle
\begin{abstract}
In this paper we present the preliminary measurement results of harvesting radio frequency(RF) energy  from the mobile phones. The aim is to revolutionize the way mobile phones are being charged and paving a way of charging the future mobile phones through RF energy harvesting. In order to measure the amount of energy that can be harvested, mobile phones from two different manufactures namely Asus and Samsung have been used. It was shown that depending on the manufacturer it is possible to harvest 1.53 joules amount of energy per day.
\end{abstract}
\IEEEpeerreviewmaketitle
\section{Introduction}
 It has been reported in \cite{numbermobile}, that the number of mobile devices used worldwide has outnumbered the total world population. The technologies are constantly evolving and demand operations that are energy-intensive. Hence, the need of the hour is to move towards more efficient sources of energy. Recently, there has been growing impetus on developing RF energy harvesting as an alternate source of energy which is readily available and is also non-polluting. 
 With ever increasing usage of mobile phone, we believe the best way to power the mobile phone can be a scenario where the mobile phone harvests energy from the radio signals emitted by other mobile phones.
 
 To explore the feasibility of harvesting RF energy from mobile phones, we measured the RF power emitted by the mobile phones when the mobile is operating in uplink where the frequency band of interest is GSM 900 MHz band. The measured RF power has been demonstrated  through the spectrum plot in Fig. 1. It can be observed from Fig. 1 that there is a spike in RF power level when a call is initiated by the mobile phone. It was also observed that the received RF power from mobile phones usually fluctuate depending on the channel conditions, distance from cellular towers etc. In this paper, we
have worked with a RF energy harvested circuit, designed
and implemented in our communication laboratory at IIIT-Delhi,
that converts the incident RF energy into DC power \footnote{For details of the circuit refer to \cite{thesisank} }.

The measurements were conducted by using mobile phones of some well-known manufacturers such as Asus and Samsung. The energy harvesting through the mobile phones has been examined on the basis of its DC output voltage. The proposed work also highlights the fact that different mobile phones have different transmit power which impacts the amount of energy harvested.

\begin{figure}
\centering
\begin{subfigure}{.5\textwidth}
  \centering
  \includegraphics[width= 70mm, height = 55mm]{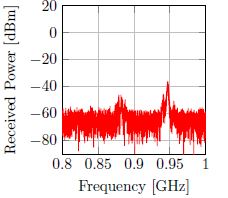}
  \caption{}
\end{subfigure}%
\vfill
\begin{subfigure}{.5\textwidth}
  \centering
  \includegraphics[width= 70mm, height = 55 mm]{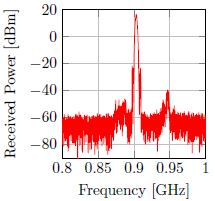}
  \caption{}
\end{subfigure}
\caption{(a) No call (b) During call at one time instant}
\label{fig: splot}
\end{figure}
\begin{figure}[t]
    \centering
\includegraphics[ width = 8.5cm, height = 5cm]{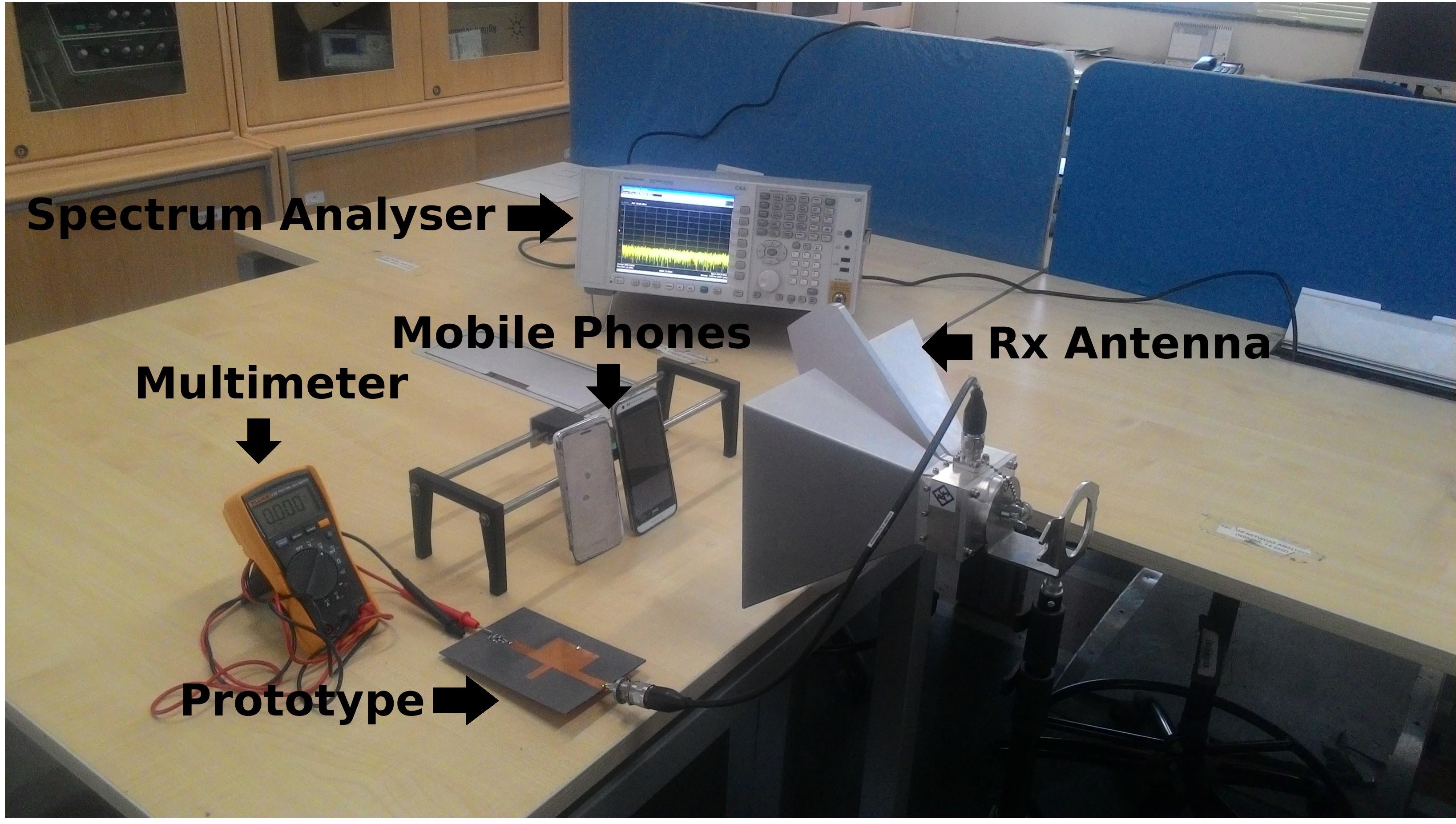}
\caption{Measurement Setup}
\label{fig: setup}
\end{figure}
   \begin{figure}
\centering
\begin{subfigure}[b]{.5\textwidth}
  \centering
  \includegraphics[width= 90mm, height = 60mm]{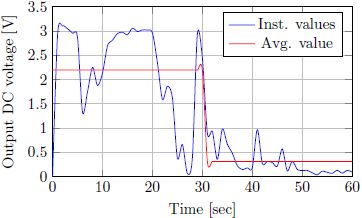}
  \caption{}
\end{subfigure}%
\vfill
\begin{subfigure}[b]{.5\textwidth}
  \centering
  \includegraphics[width= 90mm, height = 60 mm]{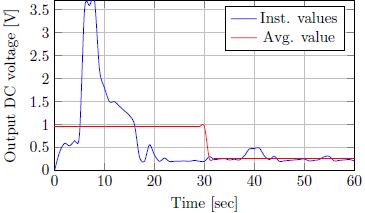}
  \caption{}
\end{subfigure}
\caption{(a) Asus Zenfone 5 Measurements (b) Samsung Galaxy Grand Max Measurements.}
\end{figure}

\section{Experimental Results}
Fig. \ref{fig: setup} exhibits the measurement setup used to measure the output DC voltage when a mobile phone is placed in the close proximity of the receiving antenna. R\&S\textsuperscript{\textregistered}HF907 \cite{105} horn antenna with a gain of upto 14 dBi has been used as a receiving antenna for the RF energy harvesting circuit. This antenna has high directivity in the frequency range of interest. As mentioned before, we have used two widely available smartphones from two well known manufacturers i.e. Asus Zenfone 5 from Asus and Samsung Galaxy Grand Max \footnote{ Interested readers can go through the specifications of this smart phones from the manufacturers' website.}. The DC output voltage of the EH circuit has been measured across a 10 k$\Omega$ load resistor using a multimeter as shown in Fig. \ref{fig: setup}. Due to imperfect soldering the load resistance reduced to a value of 8.2 k$\Omega$. For experimental validation of  the proposed setup the handsets were deliberately chosen to operate in 2G mode so that calls are placed in the desired GSM 900 MHz band. The data was collected over multiple calls from each handset. The total call duration was 1 minute which was divided into two parts : 30 seconds of call\textendash establishment phase (it includes, connection time, ringing time) and 30 seconds of call-conversation phase.

 The output DC voltage of the calling sessions has been plotted for each handset. Apart from the instantaneous values, an average DC voltage corresponding to the first 30 seconds and an average corresponding to last 30 seconds has also been plotted to demonstrate the difference in RF power levels between the two phases. Fig. 3(a) and Fig. 3(b) show the output voltage measurements for Asus Zenfone 5 and Samsung Galaxy Grand Max respectively. The voltage readings for Asus Zenfone 5 show a regular fluctuation however the average voltage for first 30 seconds is 2.2 V while for next 30 seconds is 0.4 V. For Samsung Galaxy Grand Max, the plot is almost monotonically decreasing with an average DC voltage for first 30 seconds is more than 0.5 V and next thirty seconds is approximately 80 mV and 0.26 V respectively. 
 
 One common observation from Fig. 3 is that output DC voltage has dropped by 4.9 dB between the call establishment phase and call-conversation phase. We also observed that during the call conversation phase output voltage is on a higher side when voice activity is higher as compared to low periods of voice activity. 
 
 The average DC voltage obtained over the two call phases for the Asus handset can be easily found to be :
 \begin{equation}
    \text{ Average DC voltage} = \frac{2.2 \text{ V} + 0.4 \text{ V}}{2} = 1.3 \text{ V}.
 \end{equation}
 Hence the amount of power harvested can be calculated as 
 \begin{equation}
      \text{Power harvested} = \frac{V^2}{R}= (1.3 \text{ V})\textsuperscript{2}/ (8.2 \text{ k$\Omega$}) = 0.2 \text{ mW}.
 \end{equation}
  EH circuit can be further connected to a energy accumulator, which can store energy for future use. Statistical analytics given in \cite{100} show that a person uses a smartphone on an average for 2.5 hours per day. So, the energy that can be harvested for this time duration can be determined as below 
  \begin{equation}
     \text{Energy stored per day } = \zeta \times 0.2 \text{ mW} \times 2.5 \text{ hr}
 \end{equation}
where $\zeta$ is the efficiency of the battery charger \footnote{Similarly, amount of energy stored can be calculated for other mobile phones.}. A good battery charger has an efficiency, $\zeta$ = 0.85 \cite{101}. Hence, 
\begin{equation}
     \text{Energy stored per day} = 0.85\times 0.2 \text{ mW} \times2.5 \text{ hr} = 1.53 \text{ J}
 \end{equation}
This amount of energy  can be used to operate the future mobile phones which have an inbuilt energy harvester circuit hence lowering the dependency of mobile phones on the regular batteries.

\section{Conclusion and Future Vision}
This paper explored the feasibility of self-powering the future mobiles phones through RF Energy harvesting. In this respect, measurements results were taken for two different mobile phones operating on GSM 900MHz band. Output DC voltages were measured for two call duration phases, i.e. call-establishment phase and call-conversation phase. Although the  output DC voltage fluctuates over a period of time, however we were able to harvest 0.2 mW of power or approximately 1.5 J of energy per day. 
The measurement results obtained through the designed experimental RF EH prototype shows that it is possible to harvest RF energy (albeit small amount) from the mobile phones. This motivates us to  design an EH circuit in future that can be integrated
with the mobile phones to power the mobile handset itself.

\nocite{*}
\bibliographystyle{IEEEtran}
\bibliography{main_document.bbl}
\end{document}